\documentclass{aastex701}

\newcommand{\Abacus}[1]{\textsc{#1}}
\usepackage{amsmath}
\usepackage{etoolbox}
\makeatletter
\patchcmd{\new@author@def}{http://orcid.org/}{https://orcid.org/}{}{}
\patchcmd{\new@author@def}{http://orcid.org/}{https://orcid.org/}{}{}
\patchcmd{\new@author@def}{\includegraphics[width=9pt]{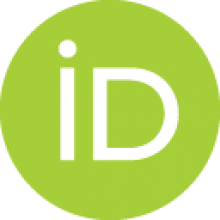}}{\includegraphics[width=12pt]{orcid-ID.png}}{}{}
\patchcmd{\new@author@def}{\includegraphics[width=9pt]{orcid-ID.png}}{\includegraphics[width=12pt]{orcid-ID.png}}{}{}
\makeatother
\defcitealias{Verza_2024_multiplicity}{V24}
\begin{document}

\title{Towards Modelling Redshift Uncertainties in the Photometric Void Size Function}

\author[orcid=0009-0005-9215-8212,gname=Pierre, sname=Boccard]{Pierre Boccard}
\affiliation{Aix-Marseille Université, CNRS/IN2P3, CPPM, 163 avenue de Luminy, Case 902, 13288 Marseille Cedex 09, France}
\email[show]{boccard@cppm.in2p3.fr}
\correspondingauthor{Pierre Boccard}

\author[orcid=0000-0002-1886-8348, gname=Giovanni, sname=Verza]{Giovanni Verza}
\affiliation{Center for Computational Astrophysics, Flatiron Institute, 162 5th Avenue, New York, NY 10010, USA}
\affiliation{Center for Cosmology and Particle Physics, Department of Physics, New York University, 726 Broadway, New York, NY 10003, USA}
\email{}

\author[orcid=0009-0001-4912-1087, gname=Giulia, sname=Degni]{Giulia Degni}
\affiliation{Aix-Marseille Université, CNRS/IN2P3, CPPM, 163 avenue de Luminy, Case 902, 13288 Marseille Cedex 09, France}
\email{}

\author[orcid=0000-0001-5620-8554, gname=Nico, sname=Schuster]{Nico Schuster}
\affiliation{Aix-Marseille Université, CNRS/IN2P3, CPPM, 163 avenue de Luminy, Case 902, 13288 Marseille Cedex 09, France}
\email{}

\author[orcid=0000-0003-2541-9433, gname=Gideon, sname=McFarland]{Gideon McFarland}
\affiliation{Department of Physics and Astronomy, Northwestern University, 2145 Sheridan Rd, Evanston, IL 60208, USA}
\affiliation{Center for Interdisciplinary Exploration and Research in Astrophysics (CIERA), Northwestern University, 1800 Sherman Ave, Evanston, IL 60201, USA}
\email{}

\author[orcid=0009-0007-3975-4079, gname=Katayoon, sname=Ghaemi]{Katayoon Ghaemi}
\affiliation{Aix-Marseille Université, CNRS/IN2P3, CPPM, 163 avenue de Luminy, Case 902, 13288 Marseille Cedex 09, France}
\email{}

\author[orcid=0000-0003-2035-9339, gname=Pauline, sname=Vielzeuf]{Pauline Vielzeuf}
\affiliation{Aix-Marseille Université, CNRS/IN2P3, CPPM, 163 avenue de Luminy, Case 902, 13288 Marseille Cedex 09, France}
\email{}

\author[orcid=0000-0002-6146-4437, gname=Alice, sname=Pisani]{Alice Pisani}
\affiliation{Aix-Marseille Université, CNRS/IN2P3, CPPM, 163 avenue de Luminy, Case 902, 13288 Marseille Cedex 09, France}
\email{}

\begin{abstract}

We present an extension of the theoretical formalism for the Void Size Function (VSF) to incorporate photometric redshift uncertainties, combining the excursion set with the theory of Lagrangian density peaks. Photometric redshift scatter induces anisotropic line-of-sight smoothing of the density field, which modifies the variance of the field and suppresses the abundance of large voids. We model this effect by damping the variance of the smoothed density field in the theoretical model, and we test it on \Abacus{AbacusSummit} halo lightcones with true redshifts and with mock photometric redshifts generated with a Gaussian scatter. The predicted damping of the variance agrees with the one measured in the halo distribution to approximately 1\% and the suppression of the VSF due to photometric redshift scatter is well reproduced by the modified model, with recalibrated barrier parameters, for $\sigma_z\leq0.015$ and for voids larger than $40$ $h^{-1}$~Mpc (about 2.5 mean tracer separations). This work is a first step towards modelling the VSF in modern photometric surveys such as the Vera C. Rubin Observatory Legacy Survey of Space and Time and the Euclid mission.
\end{abstract}

\section{Introduction}
\label{section:introduction}

The large-scale structure of the Universe contains a wealth of cosmological information to be captured by summary statistics. The statistical distribution of over-densities in the initial matter density field, the Lagrangian space, has been extensively studied and successfully modelled using robust theoretical formalisms such as the excursion set \citep{Bond_1991} or the theory of Lagrangian density peaks \citep{Bardeen_1986}. On the other hand, the statistical analysis of under-dense regions, called cosmic voids, presents a greater challenge as their formation cannot be reduced to the same simple collapse picture that describes haloes. Their evolution is instead driven by the expanding shell-crossing mechanism \citep{Sheth_2004}, and, unlike haloes, voids remain sensitive throughout their evolution to both the initial conditions and the environmental dynamics around them \citep{Schuster_2026}. In this work, we focus on the modelling of the void size function (VSF), which describes the number density of voids as a function of their radius.

The VSF measured from spectroscopic data recently provided its first ever constraints \citep{Contarini_2023}, reaching $\Omega_m = 0.29 \pm 0.06$ (about 21\% precision) and $\sigma_8 = 0.79^{+0.09}_{-0.08}$ (about 10\% precision), independent of and complementary to the constraints from standard cosmological probes. Its power from current and upcoming spectroscopic surveys is forecasted in several studies \citep{Pisani_2015, Sahl_n_2016, Sahl_n_2019, contarini_euclid_2022, Pelliciari_2023, verza_2024_Roman, song2024cosmologicalforecastvoidsize}. Nevertheless, the study of the VSF from photometric data remains limited \citep{Pollina_2019}, as a proper modelling of the VSF in photometric data is still missing. Photometric surveys have so far mostly been exploited to identify voids in projection, through two-dimensional void catalogues \citep{sanchez_2017_lensing}, rather than through a full three-dimensional treatment of the VSF such as the one we present here. The redshift uncertainty in photometric data introduces an important smearing effect along the line of sight (LOS), distorting the tracer density field and suppressing structures on all scales. While the impact of photo-$z$ uncertainties has already been studied for the galaxy power spectrum \citep{Blake_2005}, galaxy lensing tomography \citep{Ma_2006}, and galaxy angular correlations \citep{Asorey_2012}, its impact on the VSF remains unexplored, and to date no analytical model exists that predicts the VSF directly from the linear power spectrum in the presence of photo-$z$ scatter.

Our study is an extension of the theoretical formalism presented in \citet{Verza_2024_multiplicity} (hereafter V24), which combines the excursion-set framework with peak theory, for both halo and void statistics, and models statistics such as the multiplicity function, the abundance of collapsed or under-dense structures as a function of their size, and the VSF. The accuracy of the model in constraining cosmological parameters with spectroscopic data has been demonstrated in a forecast on a galaxy mock that mimics the expected features of the Roman High Latitude Wide-Area Spectroscopic Survey (HLWASS) \citep{Wang_2022} for the Nancy Grace Roman Space Telescope \citep{verza_2024_Roman}. The extension presented in our work is motivated by the advent of current and upcoming surveys producing photometric data, such as the Vera C. Rubin Observatory Legacy Survey of Space and Time \citep[LSST;][]{Ivezi__2019} and the Euclid mission \citep{Euclid_2025}. Although the extension we propose is general and can be adapted to various barrier formalisms, we focus here on its implementation in the original formalism of V24. We test it on halo lightcones from the \Abacus{AbacusSummit} suite of $N$-body simulations \citep{Maksimova_2021}. In this work, our main objective is to test whether the photometric-redshift VSF can be recovered by our extended formalism through the modification of the variance term, and to test whether our formalism predicts the amplitude and radius dependence of the suppression of the VSF at large radii $R$ in photometric data compared to the true redshift case, i.e. the case with no redshift error.
This paper is organised as follows: in Section \ref{section:basis} we review the basis of the original theoretical formalism. Section \ref{section:photoz_extension} describes the extension, introducing the anisotropic window function and the resulting modifications to the variance and barrier function. In Section \ref{section:analysis} we test our formalism on mock halo catalogues and finally make our conclusions and present future improvements in Section \ref{section:conclusion}.

\section{Basis of the formalism}

\label{section:basis}

A fundamental quantity of the excursion-set formalism is the smoothed, linearly evolved, initial density field $\delta_R$ at a comoving coordinate $\boldsymbol{q}$, where $R$ is the smoothing radius. It is defined as

\begin{equation}
    \delta_R(\boldsymbol{q}) = \int \frac{\text{d}^3k}{(2\pi)^3} \tilde{\delta}(\boldsymbol{k}) W_L(kR) e^{i\boldsymbol{k}\cdot\boldsymbol{q}},
\end{equation}

\noindent with $\tilde{\delta}$ the Fourier transform of the unsmoothed density field $\delta(\boldsymbol{q})$ and $W_L(kR)$ a spherically symmetric Lagrangian window for a real-space top hat of radius $R$

\begin{equation}
\label{eq:Wtop}
W_L(x) = 3\frac{\sin x - x\cos x}{x^3}, \qquad x\equiv kR .
\end{equation}

Following the excursion-set formalism, void formation is described by a thresholding process of the filtered Lagrangian field. For a smoothing scale $R$, if the filtered matter density field $\delta_R(\boldsymbol{q})$  crosses the negative formation threshold and has not previously crossed the positive halo formation threshold at any larger scale $r > R$, then a void is considered to be formed at the Lagrangian position $\boldsymbol{q}$. One key quantity in this formalism is the variance $S(R)$ of the smoothed field

\begin{equation}
S(R) = \sigma^2(R) = \int \frac{\text{d}^3k}{(2\pi)^3}P(k) |W_L(kR)|^2 = \frac{1}{2\pi^2} \int_0^{\infty}\text{d}k k^2 P(k)|W_L(kR)|^2,
\label{eq:Siso}
\end{equation}

\noindent with $P(k)$ being the three-dimensional matter power spectrum. The formalism uses an effective formation barrier $B(S)$, which is not a physical formation threshold, but instead encapsulates non-spherical dynamics, the features of the void finding algorithm, and the bias of the considered tracers. In the excursion-set formalism, the density contrast, which is smoothed on various scales at random positions, is described by a random walk. The barrier represents the thresholds that haloes and voids need to reach to be considered formed. Its general form is given by

\begin{equation}
    B(S) = \alpha\left[1+\left(\frac{\beta}{S}\right)^\gamma\right]
    \label{eq:B(S)} \qquad \text{and} \qquad \frac{\text{d}B}{\text{d}S} = - \alpha \beta^\gamma  \gamma S^{-\gamma-1}.
\end{equation}

In this equation, $\alpha, \beta, \gamma$ are free parameters, where $\alpha$ scales the global barrier amplitude, $\beta$ is related to the linear formation of voids, and $\gamma$ controls the shape of the moving barrier \citepalias{Verza_2024_multiplicity}. The multiplicity function $f(S)$, which is the first-crossing distribution of the forming barrier at a given scale, of the V24 model is an approximation, written in terms of the barrier value $B_S \equiv B(S)$, its derivative $B'_S \equiv \text{d}B/\text{d}S$, and the value-slope covariance combination $\Gamma_{\delta\delta}$ that we provide below. It is defined as

\begin{equation}
  \label{eq:f_up_baseline}
  f(S) \simeq \frac{e^{-B_S^2/(2S)}}{\sqrt{2\pi S}}
  \Biggl[
    \sqrt{\frac{\Gamma_{\delta\delta}}{2\pi S}}
    \exp\Biggl(
      -\frac{S}{2\Gamma_{\delta\delta}}
      B_*^{2}
    \Biggr)
    + \frac{1}{2}B_*
    \Biggl\{
      \mathrm{erf}\Biggl[
        \sqrt{\frac{S}{2\Gamma_{\delta\delta}}}
       B_*
      \Biggr] + 1
    \Biggr\}
  \Biggr],
\end{equation}

\noindent with $B_* \equiv \left(B_S/(2S)-B'_S\right)$. Here, the quantity $\Gamma_{\delta\delta}(S) \equiv S\langle\delta'^2\rangle - \langle\delta\delta'\rangle^2$ depends on the joint Gaussian statistics of the walk value $\delta(S)$ and its slope $\delta'(S)\equiv \text{d} \delta/\text{d}S$. The Lagrangian VSF relation is

\begin{equation}
\frac{\text{d}n_L}{\text{d}R} =  \frac{3}{4\pi R^3}f(R) = \frac{3}{4\pi R^3} f(S) \left|\frac{\text{d}S}{\text{d}R}\right|,
\label{eq:VSF_L}
\end{equation}

\noindent with $n_L$ being the number density of voids in Lagrangian space. The multiplicity function is a differential quantity, thus to convert the VSF into a function of Eulerian quantities, we follow the mapping $f(R)\text{d}R = f(R_E)\text{d}R_E$ where the subscript $E$ denotes Eulerian quantities throughout, which assumes that void number is conserved under the Lagrangian-to-Eulerian mapping: no voids are created, destroyed, merged, or split in the mapping from $R$ to $R_E$. The Eulerian space, also called non-linear space, is the observed space where voids are identified, as opposed to the Lagrangian space, denoting the linear initial conditions. The Eulerian multiplicity function is $f(R_E) = f(R)\,\text{d}R/\text{d}R_E = (1 + \delta_V^E)^{1/3}f(R)$, with the non-linear formation threshold being $\delta_V^E$. This parameter corresponds to the spherical non-linear evolution of the linear void formation threshold. The Eulerian radius is $R_E=R(1+\delta_V^E)^{-1/3}$ \citep{contarini_euclid_2022}, which gives the Jacobian above. We finally obtain

\begin{equation}
\frac{\text{d}n_E}{\text{d}R_E} = \frac{3}{4\pi R_E^3} f(R_E) = \frac{3}{4\pi R_E^3} (1 + \delta_V^E)^{1/3}f(S) \left|\frac{\text{d}S}{\text{d}R}\right|,
\label{eq:VSF_true}
\end{equation}

\noindent with $n_E$ being the number density of voids in Eulerian space. Note that $\left|\text{d}S/\text{d}R\right|$ here is the same Lagrangian-space derivative as in Equation~\eqref{eq:VSF_L}: since $S$ is defined as a function of the Lagrangian radius $R$ only and is never redefined in terms of $R_E$, this term is unchanged by the Eulerian mapping. Only the prefactor and the $(1+\delta_V^E)^{1/3}$ Jacobian factor differ. The resulting $\text{d}n_E/\text{d}R_E$ is the void size function (VSF) we use throughout the rest of this work.

\section{Extension with photometric redshift uncertainties}
\label{section:photoz_extension}

A simple approximation to express the photometric redshift uncertainty in cosmological surveys is to model it as a Gaussian scatter: $\sigma_z (1+z)$ \citep{Ma_2006, Chaves_Montero_2018} which, via the comoving distance-redshift relation, $\chi(z)$, with $\chi$ the comoving distance, translates to a LOS displacement with root-mean-square $\sigma_\chi$

\begin{equation}
  \sigma_\chi(z) = \left|\frac{\text{d}\chi}{\text{d}z}\right|\sigma_z(1+z)
  = \frac{c}{H(z)} \sigma_z(1+z).
  \label{eq:sigma_chi}
\end{equation}

For the \Abacus{AbacusSummit} cosmology and at $z=0.55$, the centre of the redshift range of the lightcone used in this work, we get $\sigma_\chi=17.1$, 34.1, and 51.2 $h^{-1}$~Mpc for $\sigma_z=0.005$, 0.01, and 0.015. The $\sigma_z = 0.015$ is realistic for luminous red galaxies and the lower values are used to show the effect of the scatter on the VSF. The value of $\sigma_\chi$ is almost not affected by the redshift: over our redshift range $0.3<z<0.8$ it varies by less than 3\% around its value at $z=0.55$. This stems from the factor $(1+z)$, which increases the scatter with redshift, being approximately compensated by the decrease of the Hubble distance $c/H(z)$. The observed 3D density field is the true field smeared along the LOS by a Gaussian distribution of width $\sigma_\chi$. The idea of this work is to add a Gaussian window to the original top-hat window to model the loss of LOS structure in the 3D distribution of objects. A Gaussian damping along the LOS has been used in previous studies to describe the effect of redshift errors on the power spectrum \citep{Seo_2003, Blake_2005}. This window effectively produces a Gaussian damping following the form $\exp\left(-k_\parallel^2 \sigma_\chi^2(z)/2\right)$, and in Fourier space, the new anisotropic window is

\begin{equation}
  \label{eq:Weff}
  W_{\rm ph}(\boldsymbol{k}, R)
  = W_L(kR) \exp\left(-\frac{1}{2}k_\parallel^2 \sigma_\chi^2(z)\right),
\end{equation}

\noindent where $k=|\boldsymbol{k}|$ and $k_\parallel$ is the component of $\boldsymbol{k}$ along the LOS.

From the original variance term in Equation~\eqref{eq:Siso}, we now define the photometric variance, which contains the anisotropic window. Since the variance involves $|W_{\rm ph}|^2$, the Gaussian factor of Equation~\eqref{eq:Weff} enters squared, as $\exp(-k_\parallel^2\sigma_\chi^2)$, which is the damping of the power spectrum used for photometric surveys \citep{Blake_2005}. Writing $\boldsymbol{k}=(\boldsymbol{k}_\perp,k_\parallel)$ and using the spherical change of coordinates $\mu\equiv \cos\theta = k_\parallel/k$, we obtain
\begin{align}
  S_{\rm ph}(R)
  &= \int \frac{\text{d}^3k}{(2\pi)^3} P(k)|W_L(kR)|^2 \exp\left(-k_\parallel^2 \sigma_\chi^2\right) \\
  &= \frac{1}{(2\pi)^3}\int_0^\infty \text{d}kk^2 P(k)|W_L(kR)|^2 (2\pi) \underbrace{\int_{-1}^{1}\text{d}\mu \exp\left(-k^2 \mu^2 \sigma_\chi^2\right)}_{\text{$\mu$-integral}}.
\end{align}
The $\mu$-integral becomes
\begin{align}
  \int_{-1}^{1}\text{d}\mu e^{-k^2\mu^2\sigma_\chi^2}
  &= \frac{\sqrt{\pi}}{k\sigma_\chi}\mathrm{erf}(k\sigma_\chi).
\end{align}
Substituting back, we obtain
\begin{align}
  \label{eq:S_eff_final}
  S_{\rm ph}(R)
  &= \frac{1}{4\pi^2}\int_0^\infty \text{d}kk^2 P(k)|W_L(kR)|^2 \left[\frac{\sqrt{\pi}}{k\sigma_\chi}\mathrm{erf}(k\sigma_\chi)\right] \nonumber\\
  &= \frac{1}{2\pi^2}\int_0^\infty \text{d}kk^2 P(k)|W_L(kR)|^2 G(k\sigma_\chi),
\end{align}
\noindent where we have defined $G(x) \equiv \sqrt{\pi}\,\mathrm{erf}(x)/(2x)$. The bracket in the first line of Equation~\eqref{eq:S_eff_final} is $2G(k\sigma_\chi)$, and $G\to1$ for $\sigma_\chi\to0$, which recovers Equation~\eqref{eq:Siso}. This function encapsulates the effect of the photo-$z$ smoothing on the variance. For any value of $\sigma_z > 0$, $G(k\sigma_\chi) < 1$ and thus $S_{\rm ph}(R) < S(R)$, which means that the variance of the density field is reduced at all scales due to the photo-$z$ uncertainties. Although the window $W_{\rm ph}$ is anisotropic in $k$-space, the integral over the angle between $\boldsymbol{k}$ and the LOS leaves $S_{\rm ph}$ as a function of the scalar radius $R$ only. The photo-$z$ scatter is thus described by an isotropic damping of the variance, which allows us to keep the spherical-void formalism of Section~\ref{section:basis} with $S\to S_{\rm ph}$. The derivative $\text{d} S_{\rm ph} / \text{d}R$ is straightforward and can be obtained by differentiating Equation~\eqref{eq:S_eff_final} with respect to $R$

\begin{align}
  \frac{\text{d}S_{\rm ph}}{\text{d}R}
  &= \frac{1}{\pi^2}\int_0^\infty \text{d}kk^2 P(k) W_L(kR) W_L'(kR) G(k\sigma_\chi), \qquad W_L'(kR) \equiv \frac{\text{d}W_L(kR)}{\text{d}R}.
  \label{eq:dS_eff_final}
\end{align}

In the standard formalism, the barrier $B(S)$ encapsulates the density threshold for void formation, which is determined by non-linear collapse dynamics but also depends on observational noise. Smearing the density field with photo-$z$ errors introduces observational systematics that require a recalibration of the effective barrier parameters to compensate for the loss of contrast in the density field. In the photometric case, the barrier is thus a function of the photometric variance $S_{\rm ph}$ instead of the true redshift variance $S$ and the parameters $\alpha, \beta, \gamma$ are recalibrated to fit the photometric VSF. We can write the photometric barrier as

\begin{equation}
  B_{\text{ph}}(S_{\text{ph}}) = \alpha_{\text{ph}}\left[1 + \left(\frac{\beta_{\text{ph}}}{S_{\rm ph}(R)}\right)^{\gamma_{\text{ph}}}\right].
  \label{eq:B_direct}
\end{equation}

Therefore, $\alpha_{\text{ph}}$, $\beta_{\text{ph}}$, and $\gamma_{\text{ph}}$ are different from the true redshift parameters.

Having defined the new photo-$z$ terms $S_{\rm ph}$ and $B_{\rm ph}$, we have our modified multiplicity function

\begin{align}
  \label{eq:f_up_photoz}
  f_{\rm ph}(S_{\rm ph}) \simeq &\frac{e^{-B_{\rm ph}^2/(2S_{\rm ph})}}{\sqrt{2\pi S_{\rm ph}}}
  \Biggl[
    \sqrt{\frac{\Gamma_{\delta\delta}^{\rm ph}}{2\pi S_{\rm ph}}}
    \exp\!\Biggl(
      -\frac{S_{\rm ph}}{2\Gamma_{\delta\delta}^{\rm ph}}
      (B_*^{\rm ph})^{2}
    \Biggr)
    + \frac{1}{2}B_*^{\rm ph}
    \Biggl\{
      \mathrm{erf}\!\Biggl[
        \sqrt{\frac{S_{\rm ph}}{2\Gamma_{\delta\delta}^{\rm ph}}}
        B_*^{\rm ph}
      \Biggr] + 1
    \Biggr\}
  \Biggr], \\
  &\text{with}\quad B_*^{\rm ph} \equiv \left(\frac{B_{\rm ph}}{2S_{\rm ph}}-\frac{\text{d}B_{\rm ph}}{\text{d}S_{\rm ph}}\right), \nonumber
\end{align}

\noindent where $ \Gamma_{\delta\delta}^{\rm ph}(S_{\rm ph}) \equiv S_{\rm ph}\langle\delta_{\rm ph}'^2\rangle - \langle\delta_{\rm ph}\delta_{\rm ph}'\rangle^2$. The Eulerian VSF in this case is simply

\begin{equation}
  \frac{\text{d}n_{E,{\rm ph}}}{\text{d}R_E} = \frac{3}{4\pi R_E^3} (1 + \delta_V^E)^{1/3} f_{\rm ph}(S_{\rm ph}) \left|\frac{\text{d}S_{\rm ph}}{\text{d}R}\right|.
  \label{eq:VSF_E}
\end{equation}

We expect the abundance of large voids to be suppressed in the photometric VSF compared to the true redshift one. Intuitively, the photo-$z$ scatter shuffles the tracers along the LOS. Half of the tracers are displaced by more than 11.5, 23, and 34.5 $h^{-1}$~Mpc along the line of sight (Table~\ref{tab:photoz_summary}). The photo-$z$ scattering can displace tracers from the walls of a void into its interior (or even into voids further away if the scatter is large enough), or from the interior of a void to its walls. Because the surrounding walls are, by definition, denser than the interior of a void, the net flux of tracers is into the void, which fills it and reduces its density contrast. This effect is stronger for smaller voids, which are more likely to be lost or identified at a smaller radius. For example, for $\sigma_z=0.015$, a Gaussian scatter shifts about 12\% of the tracers by more than the diameter of a void of $R=40$ $h^{-1}$~Mpc, and about 64\% for a void of $R=12$ $h^{-1}$~Mpc.
\section{Testing the model}
\label{section:analysis}
In order to validate our theoretical framework, we perform a VSF measurement on a halo mock catalogue applying a Gaussian model for the photo-$z$ scatter rather than a realistic photometric redshift simulation. This is a controlled test of the formalism itself, not yet its application to real photometric data. The main objective is to test whether the photometric redshift VSF can be recovered by our extended formalism through the modification of the variance term $S_{\rm ph}$. Suppression of the VSF at the range of radii relevant to this study (i.e. above $40$ $h^{-1}$~Mpc) is expected on physical grounds and does not by itself require a model to demonstrate; the non-trivial test is whether our formalism correctly predicts its amplitude and radius dependence, i.e. our formalism matches well with the measured VSF. We perform this test on lightcones rather than on a simulation box, as they include the radial line of sight and the redshift evolution of real data, and we also test whether the damping of the variance is necessary, by comparing with the original model without damping.
\subsection{AbacusSummit Halo Lightcones}
\label{subsection:abacus}

As mock data we use nine halo lightcones, each covering one octant of the sky, of the \Abacus{AbacusSummit} suite of $N$-body simulations \citep{Maksimova_2021, Garrison_2021}. \textsc{AbacusSummit} is one of the largest high-accuracy cosmological $N$-body datasets and is designed to meet the requirements for Stage IV surveys. The lightcones are generated with the fiducial \Abacus{AbacusSummit} cosmology given by the set of parameters \citep{Hadzhiyska_2021}: $\Omega_b h^2 = 0.02237$, $\Omega_c h^2 = 0.1200$, $h = 0.6736$, $10^9 A_s = 2.0830$, $n_s = 0.9649$, $w_0 = -1$, and $w_a = 0$. The box used to create the lightcone is 2 $h^{-1}$~Gpc\ and contains $6912^3$ particles corresponding to a particle mass of $M_{\rm part} \approx 2\times10^9\,h^{-1}M_\odot$. In this study, haloes \citep[\textsc{CompaSO};][]{Hadzhiyska_2021_CompaSO} serve as proxies for galaxies. A more realistic approach would populate them via a survey-specific Halo Occupation Distribution (HOD) model, which we leave for future work. We work in real space, i.e. without redshift-space distortions (RSD), which are not included in the model. When selecting haloes from the lightcone, we apply a mass cut of $M > 10^{13}\,h^{-1}M_\odot$, which gives us on average approximately $10^6$ haloes per catalogue, with a mean tracer separation of 15.0 $h^{-1}$~Mpc. The mass cut is chosen so that the mean tracer density, $\bar{n}\approx3\times10^{-4}~h^3\,{\rm Mpc}^{-3}$, is comparable to that of the redMaGiC luminous red galaxies \citep{Rozo_2016}, while maintaining a high enough bias to produce large voids (see below).

We focus our analysis on a redshift range of $0.3 < z < 0.8$, reducing the impact of the redshift evolution of the tracers and of voids \citep{Verza_2019,Contarini_2023,Schuster_2026}. Since the barrier parameters may have a redshift dependence \citep[see][]{verza_2024_Roman} and modelling it is beyond the scope of this work, this redshift range is a good compromise to reduce this effect while still having a large number of voids. Extending this validation to the broader redshift ranges that LSST and Euclid will probe is left for future work. We choose here to work with biased haloes, in general leading to larger voids than those identified in the cold dark matter (CDM) field at the same tracer density \citep{Sutter_2014,Pollina_2017, Contarini_2019,Schuster_2023}.

\subsection{Methodology}

\label{subsection:methodology}

To obtain a mock catalogue that resembles observed photometric data, we need to simulate the impact of photometric redshift uncertainties on the halo positions. Simulating realistic photometric redshift uncertainties is a complex task that requires a detailed understanding of the survey characteristics, the tracer population, and the photometric redshift estimation methods \citep{salvato2018flavoursphotometricredshifts, Newman_2022}. Instead we simulate the effect of photometric redshift uncertainties on the halo positions using a simple Gaussian scatter model. This is an idealised model; it captures the dominant effect of photo-$z$ errors on the LOS positions of the tracers, but it does not include non-Gaussian tails, catastrophic outliers, or dependences on galaxy type and magnitude, which are characteristic of each considered photometric survey (see Section~\ref{section:conclusion}). For each halo, we draw an observed redshift $z_{\rm ph} = z_{\rm true} + \sigma_z (1+z_{\rm true})\epsilon$, with $\epsilon$ a Gaussian random variable of zero mean and unit variance, so that $z_{\rm ph}$ has mean $z_{\rm true}$ and standard deviation $\sigma_z(1+z_{\rm true})$. We use $\sigma_z = 0.005, 0.01$, and $0.015$, to test our formalism for different levels of photometric redshift uncertainties. The photometric catalogues are built from true redshift lightcones covering $0.1<z<0.99$, and the redshift cut $0.3<z<0.8$ is applied to the scattered redshift, so that, as in real data, haloes scattered into and out of the redshift range are included. The largest value, $\sigma_z=0.015$, is similar to the scatter of the redMaGiC luminous red galaxies of the Dark Energy Survey \citep[DES;][]{DES_2016}, $\sigma_z/(1+z)\simeq0.017$ \citep{Rozo_2016}.

For both the true redshift and the different photometric catalogues, we identify cosmic voids using the updated public package Void IDentification and Examination toolkit \citep[VIDE;][]{sutter2026vide20robustscalable}, a parameter-free topological void finder algorithm based on the ZOnes Bordering On Voidness \citep[ZOBOV;][]{neyrinck_2008}. We then post-process the VIDE output following \citet{verza_2024_Roman} to obtain a catalogue whose voids satisfy the definition used in the model, i.e. a fixed mean tracer density contrast $\delta_{v,g}$ within the void, and we keep only non-overlapping voids. Additional details on the post-processing framework can be found in \citet{verza_2024_Roman}. In this work, all voids are found using a density contrast of $\delta_{v,g} = -0.6$, and we use the same threshold in the model. This value is higher than the $\delta_{v,g} = -0.7$ used in \citet{verza_2024_Roman}, which leads to larger and shallower voids. We use $\delta_{v,g} = -0.6$ as the photo-$z$ scatter fills the voids, making them shallower. We keep only voids with $R>10$ $h^{-1}$~Mpc, as smaller voids are more likely to be spurious noise fluctuations in the density field \citep{cousinou2019}.

The lightcones share the same cosmology and independent initial conditions, and we transform each of them into the three photometric catalogues (one with each $\sigma_z$), which gives 9 VSF measurements per case (36 in total), that we combine as described below. The mean number of voids per lightcone is about 4000, with some variations between the true redshifts and photometric redshift cases (Table~\ref{tab:photoz_summary}).

The data vector of our study is the Eulerian VSF, $\mathrm{d}n_E/\mathrm{d}R_E$, for the true redshifts and for each photometric case. For each of them, we sum the void counts of the 9 lightcones in each bin, $N_i$, and divide by the bin width $\Delta R_i$ and by the total comoving volume of the 9 lightcones, $V_{\rm tot}$, which is equivalent to the mean of the individual VSFs. We assign to each bin the Poisson uncertainty of the summed counts, $\sqrt{N_i}$, divided by the same normalisation, $\sigma_i=\sqrt{N_i}/(\Delta R_i V_{\rm tot})$, which is the error on the mean of the 9 VSFs, and neglect the sample variance and the correlation between bins. Previous studies have shown that the Poisson uncertainty is the dominant source of error for the VSF (\citealp{Sahl_n_2016, Contarini_2023}; \citetalias{Verza_2024_multiplicity}).

As in previous studies of the VSF \citep{contarini_euclid_2022, Contarini_2023}, we do not fit small voids: their counts are limited by the tracer sampling (small voids are lost to Poisson noise), an incompleteness that the model does not describe. These works discard post-processed void radii below 2.5 mean tracer separations. Here we discard those below about 2.5-3 times the mean halo separation of our sample, 15.0 $h^{-1}$~Mpc, i.e. $R\lesssim37.5$-45 $h^{-1}$~Mpc. Only a fraction of the voids lies in the fit windows and this fraction decreases with the photo-$z$ scatter, as the voids are moved to smaller radii: from about 12\% of the voids (about 480 per lightcone) for the true redshifts to around 1\% (about 34 per lightcone) for $\sigma_z=0.015$ (Table~\ref{tab:photoz_summary}). The statistical power of the fits therefore decreases with $\sigma_z$, and the best-fit parameters are increasingly uncertain.

\begin{deluxetable}{ccccccc}[ht!]
    \tablecaption{Line-of-sight displacements, void counts, and fit windows\label{tab:photoz_summary}}
    \tablehead{
        \colhead{Case} & \colhead{$\sigma_\chi$} & \colhead{Median $|\Delta\chi|$} & \colhead{$\langle N_{\rm voids}\rangle$} & \colhead{Fit Window} & \colhead{$N_{\rm bins}$} & \colhead{$\langle N_{\rm fit}\rangle$} \\
        & \colhead{($h^{-1}$~Mpc)} & \colhead{($h^{-1}$~Mpc)} & & \colhead{($h^{-1}$~Mpc)} & &
    }
    \startdata
        true redshift & 0 & 0 & 4112 & 40.4-86.3 & 8 & 482 \\
        $\sigma_z=0.005$ & 17.1 & 11.5 & 4450 & 41.2-75.9 & 7 & 220 \\
        $\sigma_z=0.01$ & 34.1 & 23.0 & 4496 & 39.5-63.8 & 7 & 90 \\
        $\sigma_z=0.015$ & 51.2 & 34.5 & 4421 & 38.5-53.2 & 6 & 34
    \enddata
    \tablecomments{Here, $\sigma_\chi$ is the rms line-of-sight displacement at $z=0.55$, $|\Delta\chi|$ the median line-of-sight displacement of the haloes (half of the haloes are displaced by more than this value), and $\langle N_{\rm voids}\rangle$ is the mean number of voids per lightcone. The fit window is the range of radius bins used for the fit, $N_{\rm bins}$ is the number of radius bins in the fit window, and $\langle N_{\rm fit}\rangle$ is the mean number of voids in the fit window.}
\end{deluxetable}

We fit each data vector independently with the model of Equation~\eqref{eq:VSF_true} (true redshifts) or Equation~\eqref{eq:VSF_E} (photometric redshifts), using a nested sampling method implemented in the \textsc{Dynesty} package \citep{Speagle_2020}, which allows us to explore the parameter space and obtain the best-fit parameters, as well as their uncertainties. The fit is performed with four free parameters, among which $\alpha, \beta, \gamma$ are the ones on which we focus because they describe the effective formation barrier (see Equation~\eqref{eq:B(S)}). We use flat priors, $\alpha,\beta,\mathcal{N}\sim\mathcal{U}(0,1)$ and $\gamma\sim\mathcal{U}(0,6)$. Here the cosmology is fixed to the fiducial \Abacus{AbacusSummit} cosmology as is the linear bias (see below). The fourth parameter is a normalisation constant $\mathcal{N}$ that multiplies the theoretical VSF. As in \citet{verza_2024_Roman}, it absorbs the part of the mapping from the Lagrangian prediction to the Eulerian measurement that is not captured by the excursion-set and peak-theory model, as well as the dependence on the void finder and on the tracer sampling. We treat it as an effective nuisance parameter: $\mathcal{N}<1$ means that, for the assumed barrier, the model overpredicts the number of voids found in the catalogue. We quantify the goodness of fit with the Poisson deviance of the best fit, $D=2\sum_i[\mu_i-n_i+n_i\ln(n_i/\mu_i)]$, with $n_i$ the measured and $\mu_i$ the predicted counts, which follows a $\chi^2$ distribution with $N_{\rm bins}-4$ degrees of freedom.

For the linear halo bias $b$, we use $b=2.01$, measured on the snapshot $z=0.5$, nearest available to the centre, as the large-scale ratio of the halo-matter cross-power spectrum to the matter power spectrum, for the haloes selected with the mass cut above. The bias enters the model by rescaling the variance, $S\to b^2S$: the variance used is $S_{\rm g}=b^2S$ in the true redshift model (Equation~\eqref{eq:VSF_true}) and $S_{\rm g,ph}=b^2S_{\rm ph}$ in the photometric one (Equation~\eqref{eq:VSF_E}). We use the same $b$ for all the catalogues: they contain the same haloes, with only their redshifts scattered, so the intrinsic bias is unchanged. In real data applications, the bias can, for instance, be obtained through clustering measurements of the considered tracers.

\subsection{Results and Discussion}

\label{subsection:results}

\begin{figure}[ht!]
    \centering
    \includegraphics[width=0.7\linewidth]{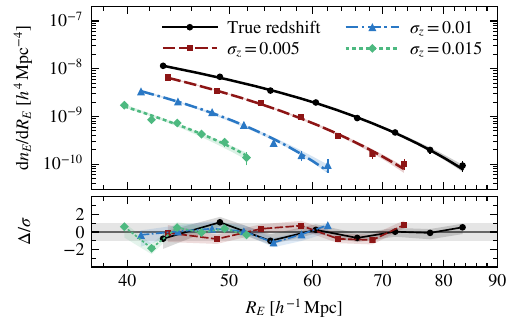}
    \caption{Comparison of the VSF in true redshifts and in photometric redshifts, measured from the \textsc{AbacusSummit} lightcones and averaged over the 9 lightcones. The cases shown are the true redshifts (black circles, solid lines) and the photometric redshifts with $\sigma_z = 0.005$ (red squares, dashed lines), $0.01$ (blue triangles, dash-dotted lines), and $0.015$ (green diamonds, dotted lines). Upper panel: points with error bars show the measurements with Poisson errors, lines the best-fit models, and shaded areas the 68\% confidence level. Lower panel: residuals of the measurements with respect to the best-fit models, in units of $\sigma$. The coloured and dark grey shaded areas show the 68\% confidence level, and the light grey horizontal band shows the $\pm 1\sigma$ interval.}
    \label{fig:vsf_comparison}
\end{figure}

The measured VSFs, averaged over the 9 lightcones, are shown in Figure~\ref{fig:vsf_comparison} for the true redshifts (black) and for $\sigma_z=0.005$, $0.01$, and $0.015$ (colours), together with the best-fit models and their 68\% confidence regions. For the true redshift and all photometric cases, the model reproduces the measured VSF within $\sim 1\sigma$ for almost all bins. The residuals of the measurements with respect to the best-fit model are shown in the lower panel of Figure \ref{fig:vsf_comparison} in units of $\sigma$.

As expected from the damping of the variance (Section~\ref{section:photoz_extension}) we observe that the photometric VSFs are suppressed at large radii with respect to the true redshift one, and the suppression increases with $\sigma_z$. The total number of voids, however, is not reduced (Table~\ref{tab:photoz_summary}): the scatter moves the voids to smaller radii, where the model does not apply, so that fewer voids lie in the fit windows (Section~\ref{subsection:methodology}). This is because, for the largest scatter that we fit, $\sigma_z=0.015$, the rms line-of-sight displacement is $\sigma_\chi\simeq51$ $h^{-1}$~Mpc, and half of the haloes are displaced by more than 34.5 $h^{-1}$~Mpc (Table~\ref{tab:photoz_summary}), which is comparable to the radii of the voids that we fit. Table~\ref{tab:barrier_parameters} gives the posterior medians and 68\% intervals of the parameters and the Poisson deviances of the best fits, which are consistent with the numbers of degrees of freedom. The normalisation is similar for the true redshifts and for the three photometric cases ($\mathcal{N}\simeq0.28$-$0.36$), and $\alpha$ and $\beta$ of the photometric cases agree with the true redshift values within their large uncertainties. We do not try to find a trend of the parameters with $\sigma_z$, since, with 6 to 8 bins, the three barrier parameters are degenerate, and the posteriors of $\beta$ and $\gamma$ are broad and multimodal for the photometric cases. The agreement of $\mathcal{N}$ among the various cases is also weak evidence that we do not focus on in this work.

\begin{deluxetable}{cccccc}[ht!]
    \renewcommand{\arraystretch}{1.5}
    \tablecaption{Posterior medians and 68\% intervals of the barrier parameters and normalisation\label{tab:barrier_parameters}}
    \tablehead{
      \colhead{Case} & \colhead{$\alpha(\alpha_{\rm ph})$} & \colhead{$\beta(\beta_{\rm ph})$} & \colhead{$\gamma(\gamma_{\rm ph})$} & \colhead{$\mathcal{N}$} & \colhead{$D/{\rm dof}$}
    }
    \startdata
      True redshift & $0.177_{-0.027}^{+0.036}$ & $0.239_{-0.213}^{+0.405}$ & $0.194_{-0.081}^{+0.096}$ & $0.290_{-0.023}^{+0.036}$ & $3.6/4$ \\
      $\sigma_z = 0.005$ & $0.250_{-0.043}^{+0.254}$ & $0.049_{-0.045}^{+0.390}$ & $0.254_{-0.178}^{+2.876}$ & $0.358_{-0.081}^{+0.082}$ & $3.4/3$ \\
      $\sigma_z = 0.01$ & $0.264_{-0.119}^{+0.292}$ & $0.130_{-0.122}^{+0.501}$ & $0.452_{-0.365}^{+4.453}$ & $0.284_{-0.142}^{+0.188}$ & $2.3/3$ \\
      $\sigma_z = 0.015$ & $0.475_{-0.273}^{+0.106}$ & $0.019_{-0.015}^{+0.553}$ & $1.161_{-1.086}^{+3.655}$ & $0.343_{-0.194}^{+0.230}$ & $3.9/2$
    \enddata
    \tablecomments{The priors are flat, and $D/{\rm dof}$ is the Poisson deviance of the best fit per degree of freedom. The photometric barrier parameters $\alpha_{\rm ph},\beta_{\rm ph},\gamma_{\rm ph}$ refer to a barrier that is a function of $S_{\rm ph}$ (Equation~\eqref{eq:B_direct}).}
\end{deluxetable}

We now ask whether the photo-$z$ damping of the variance is needed to describe the photometric VSFs. We conduct a test in which we remove the damping ($G=1$ in Equation~\eqref{eq:S_eff_final}) while leaving the barrier parameters and the normalisation free. In this case, we fit the photometric VSFs as well as with the damping, with $|\Delta\ln L|<0.1$ (Figure~\ref{fig:tests}(a)): the barrier parameters and the normalisation absorb the suppression. Do we then need the damping of the variance? We argue that we do, for three reasons.

First of all, we show in Figure~\ref{fig:damping} that our modelling of the damping of the variance is accurate. We measure the variance of the tracers in spheres of radius $R$, using 40000 spheres per realisation and radius bin for the true and photometric redshift datasets. The figure shows the ratio of the photometric variance to the true variance. We find that the measured ratios agree with the ratio between $S$ of Equation~\eqref{eq:Siso} and the prediction of Equation~\eqref{eq:S_eff_final} within approximately 1\% for the radius range $30-80$ $h^{-1}$~Mpc and for all values of $\sigma_z$. As the model describes the continuous density field, we subtract from the measured variance of the density contrast the Poisson shot noise, $1/\bar{N}$ with $\bar{N}$ the mean number of tracers in the spheres. The damping is thus the physically correct description of the effect of the photo-$z$ scatter on the field.

Secondly, letting the barrier parameters absorb the effect of the photo-$z$ scatter would make any future prediction of $\alpha$, $\beta$, and $\gamma$ harder. We already work with the assumptions that the barrier parameters encapsulate the features of the void finder and more importantly, the impact of density and bias of tracers (because these properties modify the shape of the VSF) and also probably depend on redshift, as indicated by \citet{verza_2024_Roman}. This hints at the possibility of modelling for example their redshift evolution. The future prediction of $\alpha$, $\beta$, and $\gamma$ is crucial for the application of the VSF to real data as it would allow us to fix the barrier parameters and fit only the cosmological parameters, making the constraints stronger. We also argue that the damping changes the shape of $S(R)$: at $R=50$ $h^{-1}$~Mpc, we measure $\mathrm{d}\ln S/\mathrm{d}\ln R=-2.30$ without the damping and $-2.07$, $-1.79$, and $-1.65$ for $\sigma_z=0.005$, 0.01, and 0.015. To compensate for this drop, we have to lower $\Omega_m$ (at fixed $h$) from 0.314 to 0.23, 0.17, and 0.14 respectively. We therefore suspect that the damping of the variance is also necessary to recover the correct cosmology, which we do not test here but leave for future work.

Finally, modelling the barrier parameters is not trivial: even with the damping, the barrier of the true redshift catalogue does not describe the photometric ones as shown in Figure~\ref{fig:tests}(b). For this test we fix the barrier to the true redshift best fit, with only $\mathcal{N}$ free and find that the damped model fits the photometric VSFs poorly. The damping therefore only partially captures the effect of the photo-$z$ scatter on the VSF and we argue that the barrier parameters also change with $\sigma_z$. We conclude that the photo-$z$ damping has to be encapsulated by both the damping of the variance and the barrier parameters and not only by one of them.

\section{Conclusion}
\label{section:conclusion}

In this work we have presented an extension of the theoretical formalism for the VSF for the case of photometric redshifts. We have tested it on \Abacus{AbacusSummit} halo lightcones in which the photo-$z$ uncertainties are modelled as a Gaussian scatter $\sigma_z(1+z)$, with $\sigma_z$ between 0.005 and 0.015, which is an idealised description of real error distributions. We have shown that the model is able to reproduce the suppression of the VSF: for $\sigma_z\leq0.015$, it accurately describes the photometric VSFs of voids larger than about $40$ $h^{-1}$~Mpc ($2.5$ mean tracer separations). Although we find that we can fit the photometric VSFs without the damping of the variance, we argue that it is necessary to consider it. The damping of the variance and the barrier parameters are both needed to describe the effect of the photo-$z$ scatter on the VSF.

Our analysis has several limitations. The assumption we adopt of a Gaussian scatter fixes $\sigma_z$ for all tracers, which is probably unrealistic for observational data, where it depends on the tracers. In practice, photometric surveys often encounter error distributions that deviate from a pure Gaussian distribution, exhibiting non-Gaussian tails, catastrophic outliers, systematic biases, or dependences on galaxy characteristics such as magnitude and colour \citep{Bernstein_2010, Singal_2022}. Additionally, the bias we use in the model is measured on a single snapshot, whereas the voids come from the lightcone over the redshift range $0.3<z<0.8$. Over the redshift range, $\sigma_\chi$ varies by less than 3\%, but the linear variance, which scales as the squared growth factor, varies by about $+30\%$ and $-20\%$ with respect to $z=0.55$, and the bias also evolves. We do not attempt here to model this redshift evolution, which is why we restrict the analysis to this redshift range (Section~\ref{subsection:abacus}). Finally, the model applies only to the largest voids, so that only a small fraction of the voids is used in our fits, between 1 and 12\% depending on $\sigma_z$. Moreover, for our largest scatter, comparable to the scatter of luminous red galaxies in photometric surveys, half of the tracers are displaced by more than twice the mean tracer separation and comparable to the radii of the voids we fit. Whether the model remains valid for denser samples, such as those expected from Rubin, is an open question for future work. This work also uses idealised mock catalogues. Future work will have to address its application to realistic photometric mock datasets, a more accurate description of the photo-$z$ error distributions, and a systematic characterisation of the dependence of the barrier parameters on redshift, tracer bias, and photo-$z$ scatter. Despite the mentioned limitations, this work is a first validation of the formalism for the VSF to be measured for photometric surveys, effectively opening the era of exploiting void numbers from modern photometric experiments such as Rubin or Euclid to extract cosmological constraints.

\begin{acknowledgments}
We dedicate this work to our colleague and friend Giovanni Verza. His kindness and deep knowledge will remain forever in our memories. This work was performed using the Dark Energy Center (DEC) hosted at Aix Marseille Univ, CNRS/IN2P3, CPPM, Marseille, France. PB is grateful to Julien Zoubian, Marie-Claude Cousinou, Stéphanie Escoffier, Julian Bautista, Cyrille Doux, and the LSST Dark Energy Science Collaboration (DESC) Higher-Order-Statistics team for useful discussions. PB and AP acknowledge support from the European Research Council (ERC) under the European Union’s Horizon programme (COSMOBEST ERC funded project, grant agreement 101078174). GD, KG, AP, and NS acknowledge support from the French government under the France 2030 investment plan, as part of the Initiative d’Excellence d’Aix-Marseille Université - A*MIDEX AMX-22-CEI-03. GM acknowledges support from the Center for Interdisciplinary Exploration and Research in Astrophysics Board of Visitors.
\end{acknowledgments}

\begin{contribution}
PB developed the photometric extension of the model and its implementation, performed the void identification, the measurements, and the fits, carried out the analysis, and wrote the original draft of the manuscript. The code used to compute the bias was developed using an AI tool (Claude Opus 5.5, Anthropic). AP supervised the project. GV provided expertise on the original formalism and contributed to the interpretation of the normalisation parameter. GD, NS, and PV contributed to the design of the analysis, in particular the validation tests. GM provided input on voids, LSST, and photometric redshifts, and KG on void analysis and the VSF. All authors contributed to the validation of the analysis and reviewed and edited the manuscript.
\end{contribution}

\software{VIDE \citep{sutter2026vide20robustscalable}, dynesty \citep{Speagle_2020}}

\appendix

\section{Tests with and without the damping of the variance}
\label{appendix:robustness}

\begin{figure}[ht!]
    \centering
    \gridline{\fig{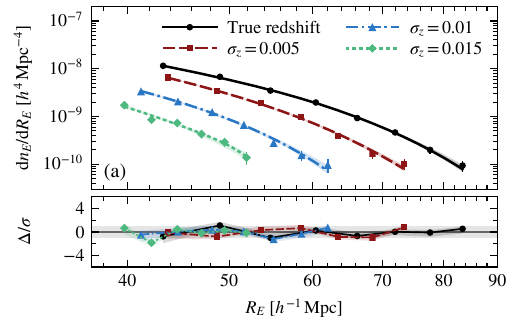}{0.49\textwidth}{}
              \fig{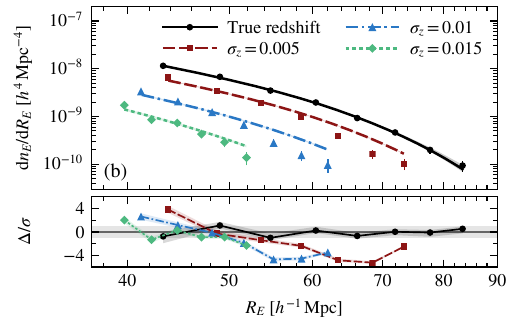}{0.49\textwidth}{}}
    \caption{Fits of the photometric VSFs for the two tests of Section~\ref{subsection:results}, with the baseline true redshift fit (black). Colours, markers, and line styles are as in Figure~\ref{fig:vsf_comparison}. Left panel (a), test A: no photo-$z$ damping ($G=1$), with free barrier parameters and normalisation. The undamped model fits as well as the damped one. Right panel (b), test B: photo-$z$ damping, with the barrier fixed to the true redshift best fit (only $\mathcal{N}$ free). The measured VSFs are steeper than the model. In both, points with error bars show the measurements with Poisson errors, lines the best fits, and shaded areas the 68\% confidence level. The lower panels show the residuals in units of $\sigma$, with the $\pm 1\sigma$ interval as a light grey horizontal band.}
    \label{fig:tests}
\end{figure}

Figure~\ref{fig:tests} shows test A in panel (a), in which the photo-$z$ damping is removed and the barrier is left free, and test B in panel (b), in which the damping is kept and the barrier is fixed to the true redshift best fit (Section~\ref{subsection:results}).

\section{Measured damping of the variance}
\label{appendix:damping}

\begin{figure}[ht!]
    \centering
    \includegraphics[width=0.8\linewidth]{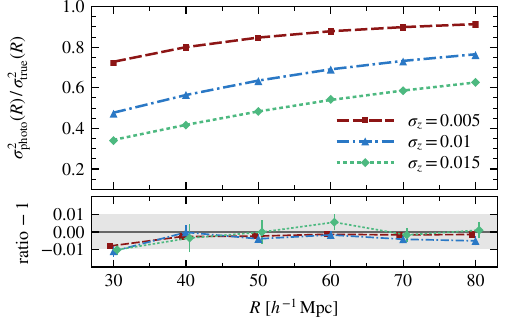}
    \caption{Ratio of the variance of the halo counts in spheres of radius $R$ between the photometric and the true redshift catalogues, for $\sigma_z = 0.005$ (red squares, dashed lines), $0.01$ (blue triangles, dash-dotted lines), and $0.015$ (green diamonds, dotted lines). Upper panel: points show the measurements (40000 random centres per radius and realisation, shot noise subtracted, mean over the 9 realisations; error bars show the standard error of the mean, i.e. the standard deviation of the 9 ratios divided by $\sqrt{9}$) and lines the model prediction $S_{\rm ph}(R)/S(R)$, with no free parameter. Lower panel: measured over predicted ratio minus one. The points are joined by lines to guide the eye and shifted by $\pm 0.5\,h^{-1}\,{\rm Mpc}$ for $\sigma_z = 0.005$ ($-$) and $0.015$ ($+$) for legibility. The grey band shows the $\pm 1\%$ interval.}
    \label{fig:damping}
\end{figure}

Figure~\ref{fig:damping} compares the damping of the variance measured in the catalogues with the model prediction (Section~\ref{subsection:results}).

\clearpage
\bibliography{biblio}{}
\bibliographystyle{aasjournalv7}

\end{document}